# A SIMPLE ANALYTICAL MODEL OF THE ANGULAR MOMENTUM TRANSFORMATION IN STRONGLY FOCUSED LIGHT BEAMS

A.Ya. Bekshaev

*I.I. Mechnikov National University, Dvorianska 2, 65082, Odessa, Ukraine*
*E-mail address*: bekshaev@onu.edu.ua

**Abstract**

A ray-optics model is proposed to describe the vector beam transformation in a strongly focusing optical system. In contrast to usual approaches basing on the focused field distribution near the focal plane, we employ the transformed beam pattern formed immediately near the exit pupil. In this cross section, details of the output field distribution are of minor physical interest but proper allowance is made for transformation of the incident beam polarization state. This enables to obtain the spin and orbital angular momentum representations which are valid everywhere in the transformed beam space. Simple analytical results are available for the transversely homogeneous circularly polarized incident beam limited only by the circular aperture. Behavior of the spin and orbital angular momenta of the output beam and their dependences on the focusing strength (aperture angle) are analyzed. The obtained analytical results are in good qualitative and reasonable quantitative agreement to the calculation performed for the spatially inhomogeneous Gaussian and Laguerre-Gaussian beams. In application to Laguerre-Gaussian beams, the model provides possibility for analyzing the angular momentum transformation in beams already possessing some mixture of the spin and orbital angular momenta. The model supplies efficient and physically transparent means for qualitative analysis of the spin-to-orbital angular momentum conversion. It can be generalized to incident beams with complicated spatial and polarization structure.

## 1. Introduction

During the last decades, light beams with inherent rotation about the propagation axis attract growing interest (see, e.g., reviews in Refs. [1–6]). In general, the rotational properties of light are expressed by the mechanical angular momentum (AM) associated with the beam electromagnetic field. The beam AM can be transmitted to other objects, e.g. microparticles [6–10], which is the most direct manifestation of the beam rotatory character. Regarding the nature and origination of the considered rotational properties, the optical AM can be subdivided in two sorts [1,11]. The spin AM (SAM) is inherent in light beams with elliptic or circular polarization and originates from the field vector rotations that take place in every point of the beam field; the orbital AM (OAM) is associated with the "macroscopic" energy circulation arising from the beam spatial inhomogeneity (for example, the well known optical vortices [1–5,12] appear as a result of helical phase distribution conjugated with the screw wavefront dislocations [2,5]). The two sorts of AM are not fully autonomous notions; moreover, only the total AM of the optical field satisfies requirements of gauge invariance while its spin and orbital parts do not [1,13]. They are well-defined quantities only in case of paraxial beams; nevertheless, in more general situations, parameters separately characterizing the spin and orbital AMs can also be introduced (for example, the spin and orbital AM fluxes [13]). Though with some theoretical imperfections, in current research practice the SAM and OAM are physically meaningful and suitable instruments of describing optical fields with rotational properties [14].

That is why the study of special features and interrelations between the two forms of the optical AM is an insistent problem of the contemporary optics. An important aspect of this problem, intensively addressed in the last few years, is the spin-to-orbital AM conversion induced by the beam transformations destroying its paraxial character, for example, sharp focusing [14–19]. Putting aside the theoretical subtleties concerning the unambiguous division of the beam angular momentum into the spin and orbital parts in the non-paraxial conditions, one can formally separate the contribution owing to the beam non-planar polarization and that originating from the beam spatial inhomogeneity, operating as follows [11,14,20]. In a monochromatic optical beam, the energy flow density (the Poynting vector time-averaged over the oscillation period) can be represented in the form

$$\mathbf{S} = \mathbf{S}_C + \mathbf{S}_O \tag{1}$$

where

$$\mathbf{S}_C = \frac{c}{16\pi k}\mathrm{Im}\left[\nabla\times\left(\mathbf{E}^*\times\mathbf{E}\right)\right], \quad \mathbf{S}_O = \frac{c}{8\pi k}\mathrm{Im}\left[\mathbf{E}^*\cdot(\nabla)\mathbf{E}\right]. \tag{2}$$

In this paper, the Gaussian system of units is used, $c$ is the light velocity, $k$ is the radiation wave number, and $\mathbf{E}$ is the complex electric field vector (the true electric field strength equals to $\mathrm{Re}\left[\mathbf{E}\exp(-i\omega t)\right]$, where the oscillation frequency $\omega = ck$). The summands $\mathbf{S}_C$ and $\mathbf{S}_O$ are the so called spin and orbital flow densities (spin and orbital currents) recently studied in detail [20–22]. Note that $\left[\mathbf{E}^*\cdot(\nabla)\mathbf{E}\right]$ is the invariant Berry notation [20] of the vector differential operation that in Cartesian coordinates reads

$$\mathbf{E}^*\cdot(\nabla)\mathbf{E} = E_x^*\nabla E_x + E_y^*\nabla E_y + E_z^*\nabla E_z$$

where, as usual, $\nabla \equiv \mathbf{e}_x\frac{\partial}{\partial x}+\mathbf{e}_y\frac{\partial}{\partial y}+\mathbf{e}_z\frac{\partial}{\partial z}$ with $\mathbf{e}_x$, $\mathbf{e}_y$, $\mathbf{e}_z$ being the unit vectors of the coordinate axes $x$, $y$, $z$. Since the Poynting vector is proportional to the electromagnetic field momentum [1,4,11], Eqs. (1) and (2) enable to express the electromagnetic angular momentum of the beam with respect to the certain reference point with radius-vector $\mathbf{R}_0$ as a sum of two terms corresponding to the summands of Eq. (1),

$$\mathbf{L} = \frac{1}{c^2}\mathrm{Im}\int\left[(\mathbf{R}-\mathbf{R}_0)\times\mathbf{S}\right]d^3R = \mathbf{L}_C + \mathbf{L}_O \tag{3}$$

which can be reduced to forms

$$\mathbf{L}_C = \frac{1}{8\pi\omega}\mathrm{Im}\int\left(\mathbf{E}^*\times\mathbf{E}\right)d^3R, \quad \mathbf{L}_O = \frac{1}{8\pi\omega}\mathrm{Im}\int(\mathbf{R}-\mathbf{R}_0)\times\left[\mathbf{E}^*\cdot(\nabla)\mathbf{E}\right]d^3R. \tag{4}$$

Here $\mathbf{R}$ is the radius vector of the current point of 3D space, the integration is performed over the whole space and it is supposed that $\mathbf{E} \to 0$ rapidly enough at $|\mathbf{R}| \to \infty$.

As is seen from Eqs. (4), the term $\mathbf{L}_C$, in contrast to $\mathbf{L}_O$, essentially involves the vector nature of the light wave and does not depend on the reference point position, which properties it shares with the SAM of a paraxial beam [1]. Moreover, in case of a paraxial beam propagating, say, along axis $z$ its expression coincides with the usual SAM definition [1,4,23], so it can be referred to as the "non-paraxial SAM". The similar but opposite arguments allow the term $\mathbf{L}_O$ to be considered as the non-paraxial OAM of a beam.

In practice, there often exists a preferential axis with respect to which the AM is defined; in most cases it is the beam propagation direction that we identify with axis $z$. Then (i) the AM longitudinal component (along axis $z$) is the main subject of interest and (ii) instead of the whole field AM given by Eqs. (3) and (4), other relevant quantities are appropriate: the AM flux through the beam cross section [13] or linear AM density (AM of a unit length of the beam taken along the propagation axis $z$) [1,4]. We will use the SAM and OAM linear densities that are defined by obvious modifications of Eqs. (3) with taking Eqs. (2) into account,

$$\mathcal{L}_C = \frac{1}{c^2} \text{Im} \int (\mathbf{r} \times \mathbf{S}_C)_z d^2 r, \quad \mathcal{L}_O = \frac{1}{c^2} \text{Im} \int (\mathbf{r} \times \mathbf{S}_O)_z d^2 r. \tag{5}$$

Here the integration is performed over the beam cross section $z$ = const, $\mathbf{r} = (x, y)$ is the transverse radius vector and index $z$ denotes the longitudinal component of a vector.

When a paraxial beam is tightly focused, its total AM (3) conserves but the initial paraxial SAM and OAM are generally redistributed between the non-paraxial SAM and OAM of the focused beam (Eqs. (4) or (5)). This effect is commonly treated as the spin-to-orbital AM conversion. The theoretical description of this phenomenon is usually based on the direct calculation of the optical field distribution $E(\mathbf{R})$ near the focal point. Regardless of what method is used, the vector Debye-type integral representation [16–18] as formulated by Richards and Wolf [24,25] or the multipole expansion of strongly focused beam [14,26,27], this way leads to bulky calculations and the representative results can only be obtained in the numerical form. In this work, we describe a simple analytical model which is probably too crude to characterize the field of the focused beam in detail but provides compact and comprehensible representation for its AM characteristics, in particular, the mutual transformations of the SAM and OAM.

## 2. Model description

Let the incident beam propagate along axis $z$ which coincides with the optical axis of the focusing system with focal distance $f$ (see Fig. 1). Let the $z$ axis origin ($z = 0$) be chosen in the location of the system exit pupil; together with the transverse coordinates $x$, $y$ it forms the Cartesian frame with unit vectors $\mathbf{e}_x$, $\mathbf{e}_y$ and $\mathbf{e}_z$. Consider a ray which approaches the exit pupil in point $N$ with coordinates $(x, y)$, or, in polar frame, $(r, \phi)$ where $r = \sqrt{x^2 + y^2}$, $\phi = \arctan(y/x)$.

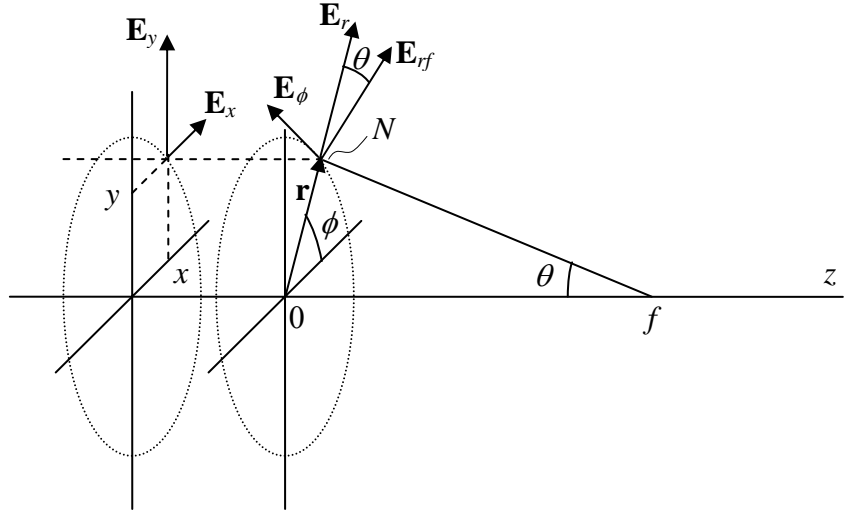

Fig. 1. Transformation of the field vectors upon focusing of a paraxial light beam (explanations in text). Vectors $\mathbf{E}_x$, $\mathbf{E}_y$, $\mathbf{E}_r$ and $\mathbf{E}_f$ are orthogonal to axis $z$, vector $\mathbf{E}_{rf}$ lies in the meridional plane $0Nf$ and is orthogonal to $Nf$.

Since the incident beam is supposed to be paraxial, this ray is parallel to axis $z$, and associated electric field contains only transverse components $E_x$ and $E_y$ (the longitudinal component $E_z$ can also be taken into consideration but because of its relatively small value [4,22] its role is inessential and will be neglected for simplicity). Upon focusing in the circularly symmetric system, each ray is bent in the meridional plane so that to cross the optical axis in the focus, i.e. it is inclined by the angle $\theta$ (see Fig. 1) which satisfies the condition

$$\cos\theta = \frac{f}{\sqrt{r^2 + f^2}} = \frac{f}{\sqrt{x^2 + y^2 + f^2}}. \tag{6}$$

Correspondingly, the polarization component lying in the meridional plane $0Nf$ is also inclined while the orthogonal component remains unchanged [14,25]. With the help of Fig. 1, one can easily see that the polarization component in the meridional plane is just the radial component $\mathbf{E}_r$, and the orthogonal one is the azimuthal component $\mathbf{E}_\phi$ so it is suitable to replace the initial representation of the polarization state in the $(x, y)$ basis by the $(r, \phi)$ basis:

$$E_r = E_x \cos\phi + E_y \sin\phi, \quad E_\phi = -E_x \sin\phi + E_y \cos\phi. \tag{7}$$

Deflection of the radial polarization component upon focusing is described by transformation of vector $\mathbf{E}_r$ into vector $\mathbf{E}_{rf}$ (see Fig. 1) with components

$$E_{rf} = E_r \cos\theta, \quad E_{zf} = E_r \sin\theta \tag{8}$$

while the azimuthal component remains unchanged: $E_{\phi f} = E_\phi$. Then, returning to the $(x, y)$ basis by reversing transformation (7), we find the Cartesian components of the electric field of the focused beam immediately after the exit pupil

$$E_{xf} = E_x \left[\cos\theta + \sin^2\phi(1-\cos\theta)\right] + E_y \sin\phi\cos\phi(\cos\theta - 1),$$
$$E_{yf} = E_x \sin\phi\cos\phi(\cos\theta - 1) + E_y \left[\cos\theta + \cos^2\phi(1-\cos\theta)\right],$$
$$E_{zf} = \left(E_x \cos\phi + E_y \sin\phi\right)\sin\theta. \tag{9}$$

Eqs. (9) are just a paraphrase of the well known rules to determine the polarization components of a strongly focused beam [25] and could also be derived from them.

Formulas (9) establish the correspondence between the incident beam and the optical field formed after the focusing system. Usually such formulas (or, rather, their equivalents that connect the input beam spatial distribution with the output field angular spectrum) play an auxiliary role: they provide initial field amplitude distributions that are substituted into the Debye integral to calculate the focused field far from the exit pupil [16–18,25]. We shall operate another way and use Eqs. (9) directly for determining the spin and orbital flows (2) and corresponding AMs (5).

In this procedure, we admit essential deviation from the real picture of the focused beam evolution. First, in our consideration the polarization transformations (9) apply to separate rays rather than to separate plane waves, and the spatial distributions of the field components in (9) follow from the ray-optics reasoning. What is more, to describe a focused beam, components (9) should be endowed with phase factors reflecting the concave wavefronts of convergent waves; and, at last, the field vector inclination presented in Fig. 1 must be accompanied by the amplitude scaling (e.g., by factor $(\cos\theta)^{-1/2}$) in order to reflect the local squeezing of the energy distribution in the inclined wave. These refinements can make the model more realistic but impede its analytical investigation. Besides, they only induce radially-symmetric field modifications and expected corrections to the energy currents (2) will most probably affect their radial components while the azimuthal ones responsible for the beam AM seem to be properly described by the "non-realistic" Eqs. (9). Of course, the field components (9) cannot be observed and even locations where Eqs. (9) are correct are not well determined but for our present purposes this is not important. During the free-space propagation, the beam AM as well as its spin and orbital parts conserve [14]. Consequently, the AM structure and its dependence on the incident beam and on the focusing parameters, calculated from Eqs. (9), are expected to provide the correct presentations of the behavior observable in the near-focus region. Also, we may hope that Eqs. (9) correctly reflect the topological peculiarities of the focused field, in particular, their wavefront singularities which are closely related to the beam OAM.

### 3. General properties and validation of the model

Once Eqs. (9) for the output field components are established, the SAM and OAM of the focused beam can be determined directly via Eqs. (2) – (5). However, before proceeding further, some important notes should be made concerning the general features of the output beam behavior.

The first important property of the considered beam transformation is revealed by comparing the magnitudes of the field vectors before and after focusing. Then Eqs. (9) give

$$|E_x|^2 + |E_y|^2 = |E_{xf}|^2 + |E_{yf}|^2 + |E_{zf}|^2. \tag{10}$$

This means that the transverse profile of the output beam energy density $I_f = (1/8\pi)|\mathbf{E}_f|^2$ exactly coincides with the incident energy density $I = (1/8\pi)|\mathbf{E}|^2$. This property is not mandatory for the transformation system, nor is it required by the energy conservation. The energy conservation is rather associated with equality of the longitudinal energy flows in the input and output fields [28], which generally does not hold for transformation (9). Then, relation (10) just accentuates the illustrative character of our model that may not reflect some secondary aspects of real focusing systems. At the same time, Eq. (13) expresses a physically meaningful feature of the developed model which, in particular, characterizes the breadth of its consistency.

Another important quantity – the vector product $\mathbf{E}_f^* \times \mathbf{E}_f$ which due to (2) determines the spin flow in the transformed beam – can be found from Eqs. (9) as

$$\mathbf{E}_f^* \times \mathbf{E}_f = i\frac{8\pi}{c} s_3(\mathbf{E}) \left(-\mathbf{e}_x \sin\theta\cos\phi - \mathbf{e}_y \sin\theta\cos\phi + \mathbf{e}_z \cos\theta\right) \tag{11}$$

where

$$s_3(\mathbf{E}) = -i\frac{c}{8\pi}\left(E_x^* E_y - E_y^* E_x\right) \tag{12}$$

is the Stokes parameter of the incident beam that characterizes the degree of circular polarization [29]. Eq. (11) certifies that the output beam SAM can only exist if $s_3(\mathbf{E}) \neq 0$, i.e. the incident beam itself possesses a certain SAM, so the "reverse" orbital-to-spin AM conversion is not possible in the discussed transformation scheme.

To make further steps, certain assumptions as to polarization state of the incident beam should be accepted. For studying the spin-to-orbital AM transformations, it is most natural to require that the incident beam be circularly polarized; for the determinacy, we consider the beam with positive helicity, so one can assume

$$E_x = E, \quad E_y = iE \tag{13}$$

where $E$ is a certain function of the transverse coordinates. In the circular polarization basis [22]

$$\mathbf{e}_\pm = \frac{1}{\sqrt{2}}\left(\mathbf{e}_x \pm i\mathbf{e}_y\right), \tag{14}$$

this incident wave is characterized by amplitudes

$$E_+ = \frac{1}{\sqrt{2}}\left(E_x - iE_y\right) = \sqrt{2}E, \quad E_- = \frac{1}{\sqrt{2}}\left(E_x + iE_y\right) = 0, \tag{15}$$

subscript "+" corresponds to the left polarization [28,29] (the spin number of a photon is +1).

Under conditions (13), Eqs. (7) – (9) for the output field components take the form

$$E_{rf} = Ee^{i\phi}\cos\theta, \quad E_{\phi f} = iEe^{i\phi}; \tag{16}$$

$$E_{xf} = Ee^{i\phi}\left(\cos\theta\cos\phi - i\sin\phi\right), \quad E_{yf} = Ee^{i\phi}\left(\cos\theta\sin\phi + i\cos\phi\right), \quad E_{zf} = Ee^{i\phi}\sin\theta. \tag{17}$$

In the circular polarization basis (14) the electric field of the transformed beam is described by the linear combination $\mathbf{E}_f = \mathbf{e}_+ E_{+f} + \mathbf{e}_- E_{-f} + \mathbf{e}_z E_{zf}$ where $E_{zf}$ is given by (17) and

$$E_{+f} = \frac{1}{\sqrt{2}} E(\cos\theta + 1) \quad E_{-f} = \frac{1}{\sqrt{2}} E e^{2i\phi}(\cos\theta - 1). \tag{18}$$

For the circularly polarized beam of Eqs. (13) or (15), the Stokes parameter (12) equals to

$$s_3(\mathbf{E}) = \frac{c}{4\pi}|E|^2 \tag{20}$$

and the quantity (11) acquires the form

$$\mathbf{E}_f^* \times \mathbf{E}_f = 2i|E|^2\left(-\mathbf{e}_x \sin\theta\cos\phi - \mathbf{e}_y \sin\theta\sin\phi + \mathbf{e}_z \cos\phi\right). \tag{19}$$

In this case, the incident beam intensity distribution $I = (c/8\pi)\cdot(2|E|^2)$ coincides with (20) and corresponding volume density of the input SAM [4] equals to

$$\mathcal{L}'_{C\text{in}} = \frac{I}{\omega c} = \frac{|E|^2}{4\pi\omega}. \tag{21}$$

Besides the polarization state, we make some natural assumptions relating the spatial distribution of the incident beam. In agreement with the common practice, we may restrict ourselves by situations where the incident beam possesses a circularly symmetric structure with possible screw wavefront dislocation:

$$E(r,\phi) = A(r)\exp(il\phi), \quad \text{Im}[A(r)] = 0. \tag{22}$$

This class of beams includes most popular beams with the optical vortices [1,2,4], for example, Laguerre-Gaussian modes (then $l$ is the integer azimuthal mode index [1,2]), which enables us to study the spin-to-orbital AM conversion not only in conditions when the incident beam carries no OAM but also when a certain amount of the initial OAM is present. We may consider only real functions $A(r)$ since the incident beam is implied to be collimated (the focusing transformation occurs at the incident beam waist cross section).

For beams satisfying condition (22), $|E|^2 = A^2$ is a radially symmetric function. Substituting it into Eq. (21) and integrating over the beam cross section, one obtains the initial SAM per unit length of the incident beam

$$\mathcal{L}_{C\text{in}} = \int \mathcal{L}'_{C\text{in}} r\,dr\,d\phi = \frac{1}{2\omega}\int_0^\infty A^2(r)r\,dr. \tag{23}$$

For analytical examples we will also use the model of a flat-top beam with constant amplitude within a circle of given radius $M$ centered at the axis $z$. This is a special case of (22) with

$$l = 0, \quad A(r) = \begin{cases} E_0, & r \geq M; \\ 0, & r \leq M. \end{cases} \tag{24}$$

This expression describes the widespread situation when a spatially homogeneous wave encounters a transversely limited circular aperture of the focusing system; the incident beam obeying Eqs.(13), (22) and (24) contains the SAM with linear density

$$\mathcal{L}_{C\text{in}} = \mathcal{L}'_{C\text{in}} \pi M^2 = \frac{1}{4\omega} E_0^2 M^2. \tag{25}$$

(cf. Eqs. (23) and (24)).

Now examine some general properties of the transformation described by Eqs. (17) and (18). The most impressive changes in the beam spatial profile occur if $E(r)$ is a regular function without phase singularities and with no OAM ($l = 0$ in Eq. (22)). First and important remark is that there appears an isotropic first-order optical vortex in the $E_{zf}$ distribution at $r = 0$, which is expressed by factor $\exp(i\phi)$ (for example see Fig. 2, bottom row). In full agreement with general properties of the optical vortices, $E_{zf} = 0$ at the vortex core $r = 0$ and its absolute value grows linearly with $r$. Emergence of this optical vortex which carries the OAM of the same handedness as the SAM of the incident beam, supplies the spectacular physical explanation for the spin-to-orbital AM conversion [14]. The same factor $\exp(i\phi)$ in equations for $E_x$ and $E_y$ is not coupled with real optical vortices because of the second multipliers in parentheses that contain helical phase distributions of the opposite sense, though anisotropic, so the whole functions possess no phase singularities. However, the top and middle rows of Fig. 2 show that the intensity and phase distributions of the transverse components experience the symmetry breakdown (possess rectangular symmetry instead of the initial circular symmetry). Since the symmetry axes of the amplitude and phase distributions do not coincide, this suggests that the non-vortex transverse components of the focused beam can also carry a sort of OAM – the so called "asymmetry" OAM [4,30].

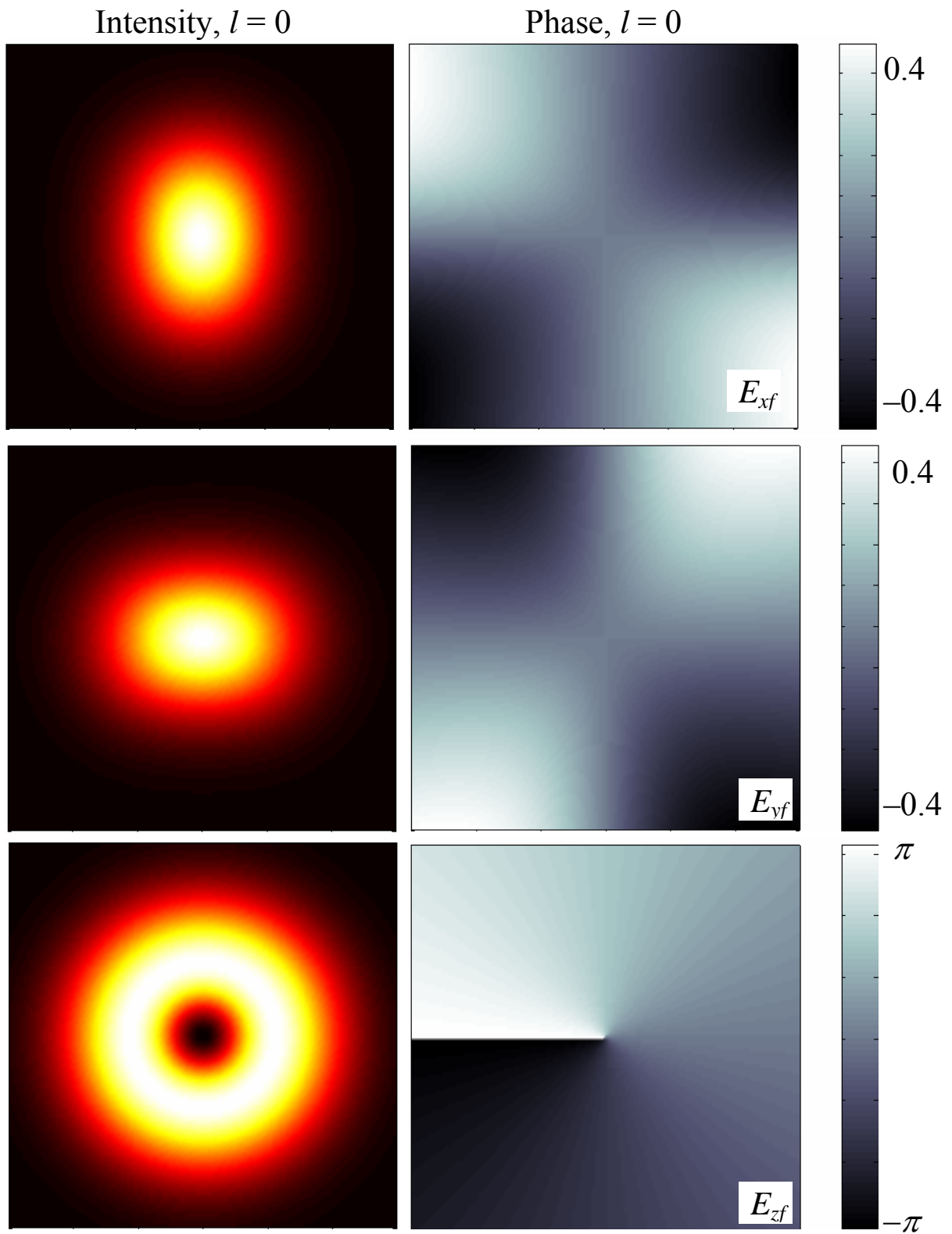

Fig. 2. Intensity (left column) and phase (right column) profiles of the transformed left-polarized Gaussian beam with initial amplitude distribution of Eqs. (13), (22) and (46) with $l = 0$, calculated in accordance with Eqs. (17) for $f/b = 0.75$ (in Eq. (47) $\sin\theta_b = 0.8$): (top row) component $E_{xf}$, (middle row) component $E_{yf}$, (bottom row) component $E_{zf}$.

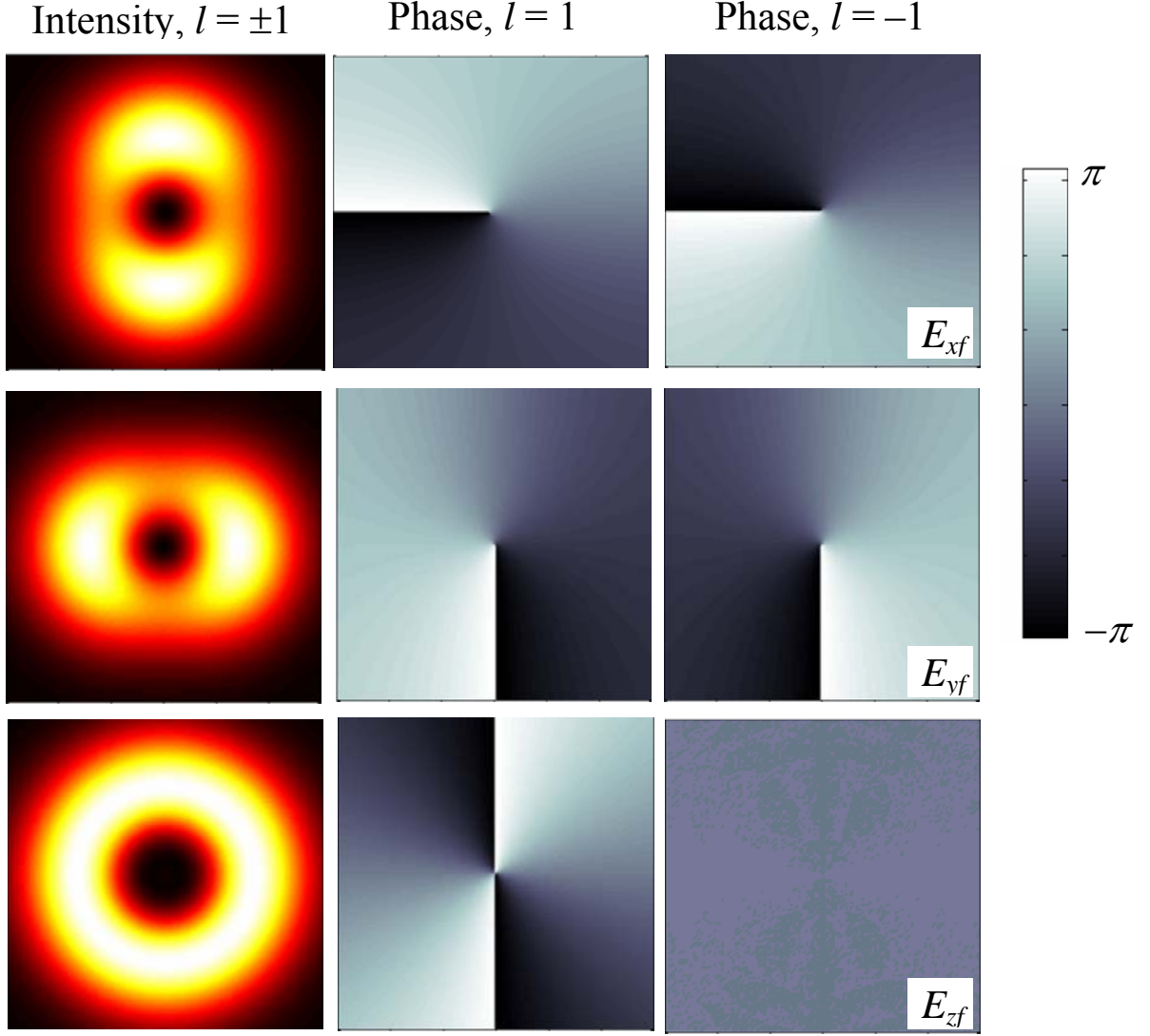

Fig. 3. Intensity and phase profiles of the transformed left-polarized Laguerre-Gaussian beam with initial amplitude distribution of Eqs. (13), (22) and (46) with $l = \pm 1$, calculated in accordance with Eqs. (17) for $f/b = 0.75$ (in Eq. (47) $\sin\theta_b = 0.8$): (top row) component $E_{xf}$, (middle row) component $E_{yf}$, (bottom row) component $E_{zf}$. Left column: intensity distributions for $l = \pm 1$ coincide, middle column: phase for $l = 1$, right column: phase for $l = -1$.

Comparison of Eqs. (15) and (18) shows that due to focusing, amplitude of the initial left-polarized component diminishes proportionally to $(\cos\theta + 1)/2 = \cos^2(\theta/2)$, and the opposite-polarized component appears with the second-order vortex of the same sense as the incident circular polarization. Note that the "double" phase singularity in this opposite component is accompanied by the near-vortex amplitude growth proportionally to $r^2$, which is seen from the Taylor expansion of the second Eq. (18) with allowance for Eq. (6).

Interestingly, when $l = 0$, the transverse profiles of the electric field components (see Fig. 2), calculated here for the cross section situated immediately after the exit pupil, look rather similar to corresponding distributions obtained by numerical simulation of the focal plane field [14,16]. By calculating instantaneous field components via $\mathrm{Re}\left[\mathbf{E}_f \exp(-i\omega t)\right]$, one can easily make sure that these instantaneous fields show features of rotation around axis $z$, which is a witness of the associated OAM [14,31]. However, for $l \ne 0$ in Eq. (22) this analogy is less impressive. The intensity profiles preserve much more features of the incident distribution and do not depend on the sign of $l$ (see left column of Fig. 3), while in the focal plane, the longitudinal component, for example, behaves quite differently at $l = -1$ and at $l = +1$ [16]. Only the wavefront topologies are reproduceed correctly. Eqs. (17) and (18) show that the $x$-, $y$- and the left-polarized components of the transformed beam preserve the initial vortex structure with the topological charge $l$ (in Fig. 3, phase vortices of the orders $\pm 1$ in the transverse components are clearly seen). The longitudinal $z$-component acquires the phase vortex with charge $l + 1$; in Fig. 3, this corresponds to the double-charged vortex for $l = 1$ and to the homogeneous phase distribution for $l = -1$ (bottom row, middle and right columns). In agreement to second Eq. (18), the emerging right polarized component carries the $(l + 2)$-order optical vortex.

Now notice that the derived equations (9) and (17) describe the field of the focused beam in a single cross section. This will permit us to evaluate the field derivatives with respect to the transverse coordinates necessary for calculation of the transverse orbital flow (second Eqs. (2) and (5)). However, for calculation of the spin flow by first Eqs. (2) and (5), $z$-derivatives of the transverse components of $\mathbf{E}_f^* \times \mathbf{E}_f$, i.e. of quantities (11), must be known. To find them, note that due to Eqs. (17) and (18) the longitudinal field $E_{zf}$ and the transverse components in the circular polarization basis $E_{+f}$ and $E_{-f}$ possess plane wavefront ($E_{+f}$) or helical wavefronts ($E_{-f}$, $E_{zf}$) with screw dislocations but no "regular" wavefront curvature. For such beams, plane $z = 0$ is the waist plane (each of the components $E_{+f}$ $E_{-f}$ and $E_{zf}$ can be expanded, say, in a series of Laguerre-Gaussian modes [2] with common waist at $z = 0$). Due to circular symmetry of the amplitude profiles, at small distances $dz$ from the waist plane the amplitude distributions vary only due to helical phase factors $\exp(i\phi)$. In the terms of Eqs. (11) or (19) these factors compensate each other, so in the first order in $dz$ these terms do not change, and the longitudinal derivatives of quantities (11) and (19) vanish. Additional arguments to support this reasoning will be given below by direct calculation of the transverse energy flows in the field described by Eqs. (16) – (18).

The inference of the above paragraph confirms that the beam characterized by Eqs. (9), (17) is not really focused (to obtain the effect of energy concentration in the focal point, proper phase factors should be imparted, cf. the note in the end of Sec. 2). However, we will shortly see that it exemplifies the main features of the spin-orbit conversion in strongly focused beams because the most important thing – tumble of the field vectors (Fig. 1) – is now taken into account in full measure. That is why we still can call the output beam of our model "focused beam".

### 4. Analysis and calculations

*4.1. Orbital angular momentum*

Now our subjects are the second Eqs. (2) and (5). Starting calculations of the orbital flow, we first remark that the longitudinal ($z$-) component of quantity $\mathbf{S}_O$ describes the main energy flow (intensity) of the beam and, in view of Eqs. (5), constitutes no interest in studying the beam AM with respect to the propagation axis. Since in the process of the spin-to-orbital AM conversion, the special role of the longitudinal field is emphasized [14], it will be suitable to represent the remaining transverse part of the orbital flow as a sum of two terms in which the transverse and longitudinal field components are separated: $\mathbf{S}_O = \mathbf{S}_{OT} + \mathbf{S}_{OL}$,

$$\mathbf{S}_{OT} = \frac{c}{8\pi k} \text{Im}\left(E_{xf}^* \,\text{grad}\, E_{xf} + E_{yf}^* \,\text{grad}\, E_{yf}\right); \tag{26}$$

$$\mathbf{S}_{OL} = \frac{c}{8\pi k} \text{Im}\left(E_{zf}^* \,\text{grad}\, E_{zf}\right) \tag{27}$$

where $\text{grad} = \mathbf{e}_x \frac{\partial}{\partial x} + \mathbf{e}_y \frac{\partial}{\partial y}$ is the symbol of the 2D transverse gradient. In further calculations it will be suitable to employ the polar coordinates in which $\text{grad} = \mathbf{e}_r \frac{\partial}{\partial r} + \mathbf{e}_\phi \frac{1}{r} \frac{\partial}{\partial \phi}$ ($\mathbf{e}_r$ and $\mathbf{e}_\phi$ are the unit vectors of the radial and azimuthal directions). Additionally, in Eq. (26) we replace the Cartesian field components $E_{xf}$, $E_{yf}$ with the polar ones $E_{rf}$, $E_{\phi f}$ by applying the reverse transformation (7) to the transformed field. Then Eqs. (26) and (27) are reduced to

$$\mathbf{S}_{OT} = \frac{c}{8\pi k} \text{Im}\left[ \mathbf{e}_r \left( E_{rf}^* \frac{\partial E_{rf}}{\partial r} + E_{\phi f}^* \frac{\partial E_{\phi f}}{\partial r} \right) + \mathbf{e}_\phi \left( E_{rf}^* \frac{\partial E_{rf}}{\partial \phi} + E_{\phi f}^* \frac{\partial E_{\phi f}}{\partial \phi} + E_{\phi f}^* E_{rf} - E_{\phi f} E_{rf}^* \right) \right], \tag{28}$$

$$\mathbf{S}_{OL} = \frac{c}{8\pi k} \text{Im}\left[ E_{zf}^* \left( \mathbf{e}_r \frac{\partial E_{zf}}{\partial r} + \mathbf{e}_\phi \frac{1}{r} \frac{\partial E_{zf}}{\partial \phi} \right) \right]. \tag{29}$$

These expressions should be evaluated with substitution of Eqs. (16) and (17). We begin with considering the radial components of (28) and (29) where the expressions in parentheses can be modified to forms

$$E_{rf}^* \frac{\partial E_{rf}}{\partial r} + E_{\phi f}^* \frac{\partial E_{\phi f}}{\partial r} = E^* \frac{\partial E}{\partial r} \left(1 + \cos^2\theta\right) + |E|^2 \cos\theta \frac{\partial}{\partial r} \cos\theta,$$

$$E_{zf}^* \frac{\partial E_{zf}}{\partial r} = E^* \frac{\partial E}{\partial r} \sin\theta + |E|^2 \frac{\partial}{\partial r} \sin\theta.$$

For considered beams satisfying the condition (22) these expressions are real and so they give zero contributions to the flows (28), (29). Consequently, the whole orbital flow possesses no radial component. The fact that, in a beam with radially symmetric profile, the transverse energy flow possesses no radial component means that the beam transverse intensity profile does not change in the first order upon small longitudinal displacement of the observation plane (e.g., from $z = 0$ to $z = dz$). Therefore, z-derivatives of the x- and y-components of $\text{Im}\left(\mathbf{E}_f^* \times \mathbf{E}_f\right)$ in Eq. (19), which are proportional to $|E|^2$, really vanish, and this is an additional confirmation of possibility to omit the z-derivatives in the spin flow calculation, which was substantiated in the last paragraphs of Sec. 3.

Azimuthal components of (28) and (29) are easily derived with allowance for (22), and after simple calculations one obtains

$$\mathbf{S}_{OT} = \frac{\mathbf{e}_\phi}{r} \frac{c}{8\pi k} A^2 \left[-2\cos\theta + (l+1)\left(1 + \cos^2\theta\right)\right], \quad \mathbf{S}_{OL} = \frac{\mathbf{e}_\phi}{r} \frac{c}{8\pi k} A^2 (l+1) \sin^2\theta \tag{30}$$

and

$$\mathbf{S}_O = \mathbf{S}_{OT} + \mathbf{S}_{OL} = \frac{\mathbf{e}_\phi}{r} \frac{c}{4\pi k} A^2 (l + 1 - \cos\theta). \tag{31}$$

One can easily see that if $\cos\theta \to 1$, or, in accordance with (6), $f \to \infty$ (the focusing action disappears), the total orbital flow reduces to $\frac{\mathbf{e}_\phi}{r} \frac{c}{4\pi k} lA^2$ – the orbital flow of the incident beam described by Eqs. (22). The difference between (31) and this value just represents the net effect of the spin-to-orbital flow conversion

$$\Delta \mathbf{S}_O = \frac{\mathbf{e}_\phi}{r} \frac{c}{4\pi k} A^2 (1 - \cos\theta). \tag{32}$$

The OAM of the focused beam follows from Eqs. (28), (29) and the second Eq. (5). It is convenient to express the results for the OAM and its separate constituents in units of the initial SAM (23) carried by the incident beam; then Eq. (5) due to azimuthal direction of $\mathbf{S}_O$ leads to representation

$$\Lambda_O = \frac{\mathcal{L}_O}{\mathcal{L}_{C\,\text{in}}} = \frac{1}{c^2} \frac{2\pi}{\mathcal{L}_{C\,\text{in}}} \int_0^\infty r |\mathbf{S}_O(r)| r dr. \tag{33}$$

From now on, we will denote such dimensionless quantities by Greek $\Lambda$ with corresponding indices, keeping the Euclid letter $\mathcal{L}$ for the absolute AM measures.

For any real beam, convergence of the integral in (33) is ensured by the limited transverse size of the beam but, in general, the integral cannot be evaluated in closed form. To proceed the calculations further, a certain form of the radial amplitude distribution $A(r)$ should be specified. Analytical results are available for the spatially homogeneous incident beam of Eq. (24). In this case, the orbital flow constituents are determined by Eqs. (30) with $l = 0$; after they are integrated similarly to Eq. (33) and normalized by the input beam SAM (25), with taking Eq. (6) into account we arrive at

$$\Lambda_{OT} = \frac{1}{M^2} \int_0^M \left(1 - \frac{f}{\sqrt{r^2 + f^2}}\right)^2 r dr = \frac{1}{2}\left(1 + \tau^2 \ln\frac{1+\tau^2}{\tau^2} - 4\tau\sqrt{1+\tau^2} + 4\tau^2\right)$$

$$= \frac{1}{2}\left(1 - \cot^2\theta_M \ln\cos^2\theta_M\right) - 2\frac{\cos\theta_M (1 - \cos\theta_M)}{\sin^2\theta_M}, \tag{34}$$

$$\Lambda_{OL} = \frac{1}{M^2} \int_0^M \frac{r^3}{r^2 + f^2} dr = \frac{1}{2}\left(1 - \tau^2 \ln\frac{1+\tau^2}{\tau^2}\right) = \frac{1}{2}\left(1 + \cot\theta_M \ln\cos^2\theta_M\right) \tag{35}$$

and

$$\Lambda_O = \Lambda_{OT} + \Lambda_{OL} = 1 - 2\tau\sqrt{1+\tau^2} + 2\tau^2 = 1 - 2\frac{\cos\theta_M (1 - \cos\theta_M)}{\sin^2\theta_M}. \tag{36}$$

where

$$\tau = \frac{f}{M} = \cot\theta_M, \tag{37}$$

$\theta_M$ is the aperture angle. Expression (36) vanishes when $\tau \to \infty$ ($\theta_M \to 0$), i.e. when no focusing takes place; its non-zero value at finite $\tau$ just describes the sought effect of the transformation of the initial SAM of the incident beam into the OAM of the focused beam.

*4.2. Spin angular momentum*

For the incident beam with circular polarization (13) the SAM is calculated directly by substituting expression (19) into first Eq. (2). Discarding derivatives with respect to $z$ in agreement with the last note of Sec. 3 and the discussion following Eq. (29), we obtain the spin flow density

$$\mathbf{S}_C = \mathbf{e}_x S_{Cx} + \mathbf{e}_y S_{Cy} = \frac{c}{8\pi k}\left[\mathbf{e}_x \frac{\partial}{\partial y}\left(|E|^2 \cos\theta\right) - \mathbf{e}_y \frac{\partial}{\partial x}\left(|E|^2 \cos\theta\right)\right]. \tag{38}$$

Quite expectedly, $S_{Cz} = 0$ and the whole spin flow is of transverse character. In Eq. (38) the dependence on $x$ and $y$ is contained in $|E|^2$ and in $\cos\theta$ (see Eq. (6)). If the incident beam satisfies Eq. (22), the quantity in parentheses of (38), $A^2\cos\theta$, depends only on $r$, and by employing formulas

$$\frac{\partial}{\partial x} = \frac{x}{r}\frac{\partial}{\partial r}, \quad \frac{\partial}{\partial y} = \frac{y}{r}\frac{\partial}{\partial r}$$

and Eq. (6), expression (38) can be transformed to

$$\mathbf{S}_C = \frac{c}{8\pi k} \frac{\mathbf{e}_x y - \mathbf{e}_y x}{r} \frac{\partial}{\partial r}\left(A^2 \cos\theta\right) = \frac{c}{8\pi k} \mathbf{e}_\phi \frac{\partial}{\partial r}\left(A^2 \cos\theta\right) = \frac{c}{8\pi k} \mathbf{e}_\phi \cos\theta \left(A^2 \frac{r}{r^2+f^2} - \frac{\partial A^2}{\partial r}\right). \quad (39)$$

Note that in full agreement with the spin flow nature due to which it is always directed along the constant-level lines of function $s_3(x, y)$ [21,22] – now circumferences centered at axis $z$ – the radial component of $\mathbf{S}_C$ vanishes. For the homogeneous incident beam of Eq. (24), Eq. (39) can be simplified to

$$\mathbf{S}_C = \mathbf{e}_\phi \frac{c}{8\pi k} E_0^2 \frac{fr}{\left(r^2+f^2\right)^{3/2}}. \quad (40)$$

Expression (40) tends to zero when $f \to 0$ ($\cos\theta \to 1$) which means that at extremely strong focusing the spin flow is likely to completely convert into the orbital form of the transverse energy circulation. However, it also disappears when $f \to \infty$ (no focusing) which is in compliance with the fact that in spatially homogeneous beams with circular polarization the macroscopic spin flow is absent [21,32].

The SAM associated with this flow is calculated by means of first Eq. (5). Analytical results, again, will be obtained for the homogeneous incident beam satisfying Eqs. (22) and (24). Then the first Eq. (5) and Eq. (25) give

$$\Lambda_{CI} = \frac{\mathcal{L}_{CI}}{\mathcal{L}_{C\,\text{in}}} = \frac{1}{8\pi\omega} \frac{1}{\mathcal{L}_{C\,\text{in}}} \int |\mathbf{r} \times \mathbf{S}_C| r dr d\phi$$

$$= \frac{E_0^2 f}{4\omega} \frac{1}{\mathcal{L}_{C\,\text{in}}} \int_0^M \frac{r^3 dr}{\left(r^2+f^2\right)^{3/2}} = \tau\left(\sqrt{1+\tau^2} + \frac{\tau^2}{\sqrt{1+\tau^2}} - 2\tau\right) = \frac{\cos\theta_M \left(1 - \cos\theta_M\right)^2}{\sin^2\theta_M}. \quad (41)$$

This expression represents the SAM associated with the macroscopic spin flow inside the considered cross section area. However, for beams with abrupt transverse boundary this is only a part of the total SAM [4,32]. The full SAM carried by such a beam includes also the boundary contribution which is determined as

$$\Lambda_{CB} = \frac{1}{16\pi\omega} \frac{1}{\mathcal{L}_{C\,\text{in}}} \oint_M \text{Im}\left(\mathbf{E}^* \times \mathbf{E}\right)_z |\mathbf{r} \times d\mathbf{r}| = \frac{1}{2\pi M^2} \oint_M \cos\theta |\mathbf{r} \times d\mathbf{r}| = \frac{\tau}{\sqrt{1+\tau^2}} = \cos\theta_M; \quad (42)$$

here the integral has been taken along the circumference bounding the beam cross section. Adding both contributions provides the whole SAM of the transformed beam

$$\Lambda_C = \Lambda_{CI} + \Lambda_{CB} = 2\tau\left(\sqrt{1+\tau^2} - \tau\right) = 2\frac{\cos\theta_M \left(1 - \cos\theta_M\right)}{\sin^2\theta_M}. \quad (43)$$

It is instructive to verify this result by the elementary estimates, following to the way described in Ref. [14]. Due to (21), before the focusing, an element $dw$ of the beam cross section carries the SAM $\mathcal{L}'_{C\,\text{in}} dw$; after the focusing, its longitudinal component reduces to $\mathcal{L}'_{C\,\text{in}} dw \cos\theta$ (see Fig. 1). The whole SAM of the focused beam can thus be found by integrating this quantity over the whole beam cross section,

$$\frac{E_0^2}{4\pi\omega} \frac{1}{\mathcal{L}_{C\,\text{in}}} \int \cos\theta\, r dr d\phi = \frac{2f}{M^2} \int_0^M \frac{r dr}{\sqrt{r^2+f^2}} = 2\tau\left(\sqrt{1+\tau^2} - \tau\right), \quad (44)$$

which exactly coincides with (43). This serves an additional witness to the relevance of the accepted model. We can also remark that when the focusing action disappears (at $\tau \to \infty$, $\theta_M \to 0$), results (43) and (44) tend to 1, i.e. the output beam SAM quite expectedly reduces to the SAM of the unperturbed incident beam. In all other cases, the resulting SAM of the output beam (43) or (44) is less than the incident beam AM, which implies that a part of the initial SAM is transformed to the

orbital one. That this is really so, follows directly from confronting Eqs.(43) and (36), which readily gives

$$\Lambda = \Lambda_O + \Lambda_C = 1, \qquad (45)$$

in agreement with the AM conservation requirements.

### 5. Results and discussion

The main importance of Eqs. (34) – (36) and (41) – (43) is that they enable to study the process of the spin-to-orbital AM conversion in a simple analytical manner, with explicit exhibition of involved factors and their interaction. As an example, the main regularities of the AM transformation upon focusing the spatially homogeneous incident beam, that satisfies Eqs. (13), (22) and (24), are illustrated by Fig. 4. It presents how the AM related quantities (34) – (36) and (41) – (43) depend on the aperture angle $\theta_M$ defined by Eq. (37). It is seen that, indeed, the OAM $\Lambda_{OL}$ associated with longitudinal field $E_{zf}$ dominates; however, the contribution associated with the transverse field components $\Lambda_{OT}$ also exists and rapidly grows with the focusing strength so that it becomes commensurable with $\Lambda_{OL}$ at very strong focusing. In accordance with reasoning of Sec. 3, below Eq. (25), the $\Lambda_{OL}$ can be associated with the vortex-type OAM while the $\Lambda_{OT}$ represents the asymmetry OAM component [30] of the focused beam. In the limit $\theta_M \to 90°$ both OAM contributions appears to be equal and their sum reaches the initial SAM value. Apparently, the whole SAM tends to zero, i.e. is fully converted to the OAM. The conclusion about the 100% AM conversion may seem too rigorous and contradicts to more accurate calculations [14,17,18]. This controversy can be attributed to the model approximations; besides, the focusing angles close to 90°, where the predicted effect is anticipated, do not seem to be physically realizable.

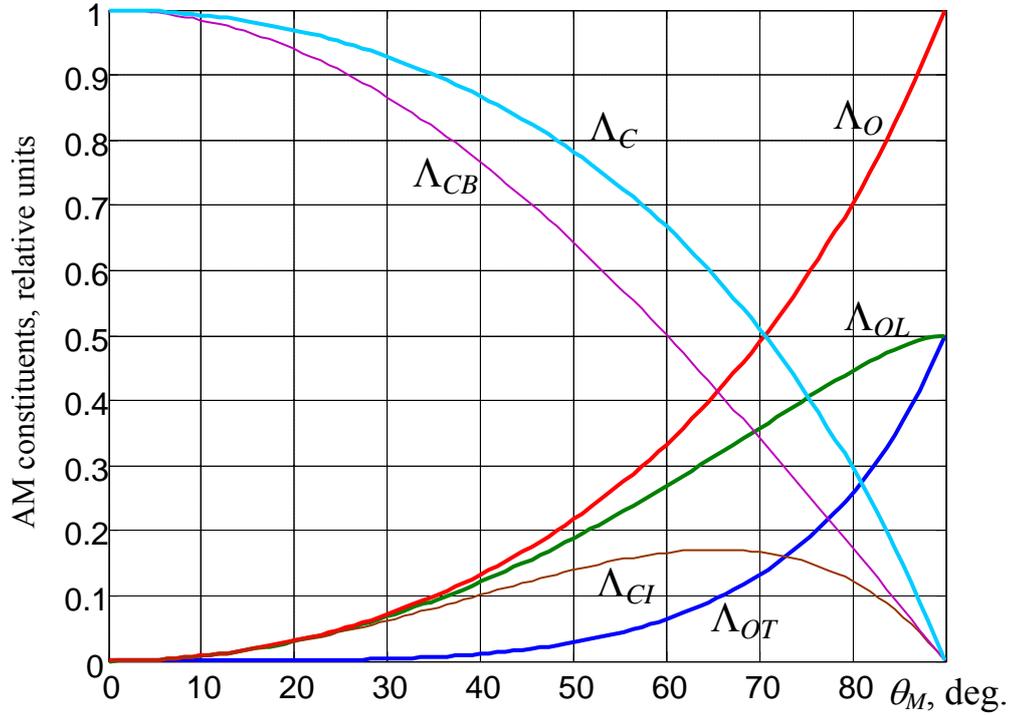

Fig. 4. AM constituents of the focused spatially homogeneous circularly polarized beam vs aperture angle; each curve is marked by the corresponding quantity notation from Eqs. (34) – (36) and (41) – (43).

As could be expected, almost the whole SAM of the focused beam is associated with the beam boundary (curve $\Lambda_{CB}$ lies higher than curve $\Lambda_{CI}$) because the beam amplitude distribution is close to homogeneous and does not vanish at the beam transverse boundary (see, e.g., Eqs. (17)). With stronger focusing, the output beam spatial inhomogeneity becomes more essential, and contributions $\Lambda_{CB}$ and $\Lambda_{CI}$ approach together but practical importance of this effect is minor because the whole SAM in these conditions is very small. The whole beam AM always obeys the conservation law (45).

However, the value of the developed approach extends far beyond the possibility to analyze the transformation of the spatially homogeneous beam of Eq. (24). It can be easily applied to more realistic spatially inhomogeneous beams. In this case analytical formulas for the AM constituents (34) – (36) and (41) – (43) used in Fig. 2 are not valid but equations describing the transformed field (9), (11) and (16) – (20) are still correct. So the spin flow can be computed directly from Eq. (38) and the orbital one follows from Eqs. (26) – (29). Then $\mathcal{L}_{CI}$ can be calculated by substitution of $\mathbf{S}_C$ into the first Eq. (5) and $\mathcal{L}_{OT}$, $\mathcal{L}_{OL}$ similarly ensue from $\mathbf{S}_{OT}$, $\mathbf{S}_{OL}$ and the second Eq. (5). The boundary part of the SAM can be determined via the first Eq. (42) where the contour of integration should be replaced with boundary of the domain used for numerical integration in first Eq. (5). Dimensionless quantities $\Lambda_C$, $\Lambda_{OT}$ and $\Lambda_{OL}$ are formed with employment of normalization integral (23).

The results for the simplest representatives of the Laguerre-Gaussian family are displayed in Fig. 5. The incident beams are supposed circularly polarized (satisfy Eqs. (13) or (15)) with the initial amplitude distribution of Eq. (22) with

$$A(r) = Cr^{|l|} \exp\left(-\frac{r^2}{2b^2}\right) \tag{46}$$

where $C$ is inessential normalization constant; behavior of the output beam AM characteristics is studied in their dependence on the effective aperture angle

$$\theta_b = \operatorname{arc\,cot}(f/b). \tag{47}$$

A general comment to Fig. 5 is that due to transverse confinement of the involved beams, the beam amplitude at the integration domain boundary practically vanishes together with the associated SAM contribution $\Lambda_{CB}$, and the whole SAM of the focused beam is thus determined by $\Lambda_{CI}$. With allowance for this inessential remark, direct comparison of Fig. 5a and Fig. 4 witnesses that the case of Gaussian beam differs only by some quantitative variations from the case of spatially homogeneous beam. This allow us to hope that analytical formulas of the above section can be used, at least as a first approximation, for description of the spin-to-orbital AM transformations for various incident beams with wide range of spatial configurations. Herewith, it may be necessary to find an appropriate equivalent of the aperture angle (37): it is a great deal of fortune that for the Gaussian beam (46), the "natural" choice (47) provides apparently good result.

When $l \neq 0$, Eqs. (22) and (46) provide examples of the AM transformation for beams initially possessing both the SAM and OAM. The simplest cases of $l = 1$ and $l = -1$ are illustrated by Figs. 5b, c. In both cases, focusing leads to transformation of the initial SAM into the orbital one, which is added to the initial OAM of the beam. In this case, the transverse-field constituents of the OAM (curves $\Lambda_{OT}$) are no longer associated with the asymmetry part of the OAM, since the incident beam carries the OAM which belongs to the transverse field components and is of the vortex character [4,30]. If the initial SAM and OAM are of the same sense, Eqs. (17), (22) and (46) show that the transverse field components of the focused beam carry the first-order optical vortices while the vortex in $E_{zf}$ is double-charged. The OAM absolute value grows due to focusing (Fig. 5b), and ultimately the whole beam AM becomes of the orbital nature.

When the initial SAM and OAM are opposite ($l = -1$, Fig. 5c), analogous reasoning show that components $E_{xf}$, $E_{yf}$ in (17) possess vortices of charge $-1$ and $E_{zf}$ looses the phase singularity. The longitudinal field contribution to the output beam OAM vanishes (curve $\Lambda_{OL}$ coincides with the horizontal axis) and the whole OAM of the focused beam is associated with the transverse field components (curves $\Lambda_O$ and $\Lambda_{OT}$ are the same). Of course, the OAM arising from the converted SAM is opposite to the initial OAM and compensates it up to full vanishing at $\theta_b \to 90°$. The total beam AM is always constant in all cases ($\Lambda = 1$ in Fig. 5a, $\Lambda = 2$ in Fig. 5b and $\Lambda = 0$ in Fig. 5c).

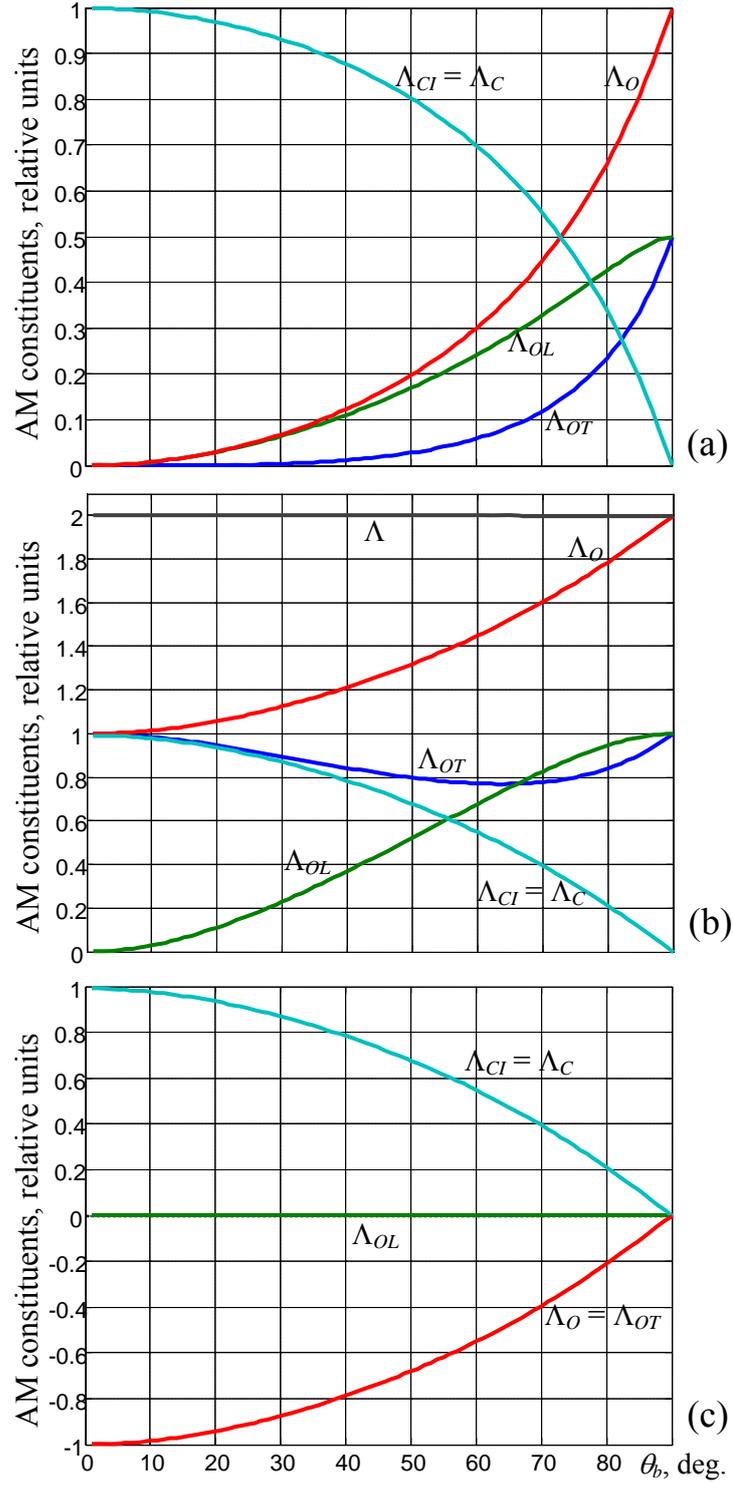

Fig. 5. Behavior of the AM constituents upon focusing the left-polarized beams of Eq. (13) with initial spatial inhomogeneity of Eqs. (22), (46): (a) Gaussian beam ($l = 0$), (b) Laguerre-Gaussian beam with equal spin and orbital AMs ($l = 1$) and (c) Laguerre-Gaussian beam with initially opposite spin and orbital AMs ($l = -1$). Each curve is marked by the corresponding quantity notation, $\theta_b$ is given by Eq. (47).

## 6. Conclusion

The model of the spin-to orbital AM conversion in strongly focused beams, presented in this paper, allows to describe, in a simple analytical manner, the main regularities of the conversion process, its dependence on the conversion parameters (focal length, aperture angle) and provides clear physical interpretation of the results obtained. Despite that explicit analytical formulas are only available for the case of spatially homogeneous incident beam, their value is noticeably wider. Numerical calculations for the beam with Gaussian transverse intensity distribution have shown that the simple analytical results obtained for the flat-top beam provide perfect qualitative and reasonable quantitative description of the spin-to-orbital conversion of spatially inhomogeneous non-vortex beams.

Due to the model simplicity, it retains the physical transparency even in cases where the numerical calculations are indispensable. The model can be easily generalized to admit the elliptically polarized and asymmetric incident beams: in such cases, Eqs. (13) and (22) should be replaced but the crucial equations (9) – (12), (26) – (29) and (38) are still correct and can be used for evaluation of the spin and orbital flows with subsequent calculation of the SAM and OAM by means of Eqs. (5).

However, the model limitations should also be recognized. The model is essentially of the geometric-optics character: we derived the transformation rules (9) for a ray, and apparently plane-wave plane parameters of (9) explicitly depend on the transverse coordinate via $\theta$. Therefore, diffraction effects appear outside the consideration, in particular, we do not account for the edge diffraction even in case of the flat-top incident beam of Eq. (24). Another important note is that the transformed field of our model (for example, what is described by Eqs. (9)) is, in fact, imaginary one and can hardly exist somewhere in the real physical space. At the same time, the SAM and OAM calculated from this field are expected to be much closer to reality (see the notes at the end of Sec. 2).

All the above drawbacks are inherent in the model and cannot be removed without destroying its main advantages: mathematical simplicity and physical transparency. Hence, it will hardly be acceptable for accurate computations and quantitative experimental data analyses. Nevertheless, it can be relevant and helpful for qualitative study of the spin-to-orbital transformations of beams with complex spatial or polarization structure and for the physical interpretation of the spin-to-orbital AM conversion in strongly focused beams, e.g. in the teaching practice.